# Sticky information and price controls: Evidence from a natural experiment

Doron Sayag [a,b], Avichai Snir [a], Daniel Levy [a,c,d,e,f] *

[a] *Department of Economics, Bar-Ilan University, Ramat-Gan 5290002, Israel*
[b] *The Office of National Statistician, Central Bureau of Statistics, Jerusalem 9546456, Israel*
[c] *Department of Economics, Emory University, Atlanta, GA 30322, USA*
[d] *International Centre for Economic Analysis, Wilfrid Laurier University, Waterloo, Ontario, Canada*
[e] *International School of Economics at Tbilisi State University, 0108 Tbilisi, Georgia*
[f] *The Rimini Centre for Economic Analysis (RCEA)*

**Abstract:** We test the predictions of the sticky information model using a survey dataset by comparing shoppers' accuracy in recalling the prices of regulated and comparable unregulated products. Because regulated product prices are capped, they are sold more than comparable unregulated products, while their prices change less frequently and vary less across stores and between brands, than the prices of comparable unregulated products. Therefore, shoppers would be expected to recall the regulated product prices more accurately. However, we find that shoppers are better at recalling the prices of unregulated products, in line with the sticky information model which predicts that shoppers will be more attentive to prices that change more frequently.

* Corresponding author at: Department of Economics, Bar-Ilan University, Ramat-Gan 5290002, Israel.
  *E-mail address:* Daniel.Levy@biu.ac.il (D. Levy)

## 1. Introduction

Sticky information models assume that information processing is costly. Therefore, agents may rationally ignore some information, depending on its costs and benefits.[1] We contribute to the literature by taking advantage of a unique regulatory setting to test the prediction of the sticky information model using micro-level data. As Reis (2006b) notes, consumers incur information-gathering and processing costs. Consequently, "…they rationally choose to only sporadically update their information and re-compute their optimal consumption plans. In between updating dates, they remain inattentive" (Reis, 2006b, p. 1761).

We study consumers' price awareness for two groups of regularly purchased food products. The prices of products in the first group are subject to price control regulation ("*regulated products*"), constraining retailers from setting prices that exceed a cap price. The products in the second group are comparable to the first group but are set without regulators' intervention ("*unregulated products*").

The sticky information model predicts that consumers will update their information regarding the prices of regulated products less often than the prices of comparable but unregulated products (Dickson and Sawyer 1990, Urbany et al. 1996, Reis 2006b) because regulated goods' prices change less frequently than the prices of unregulated goods (Dexter et al. 1993 and 2002).

We test this hypothesis using a survey, where we asked shoppers coming out of supermarkets, to recall the prices of regulated products, and of comparable unregulated products. We find that consumers are less accurate in recalling the prices of regulated products in comparison to the prices of comparable unregulated products.

In section 2, we briefly describe the price controls. In section 3, we present the data. In section 4, we discuss the estimation results. In section 5, we discuss the robustness of our findings and conclude. In the Online Supplementary Web Appendix, we present the details of the robustness tests we ran.

## 2. Background: Price controls in Israel

In Israel, price controls have been in place since July 1985, adopted as part of an inflation stabilization program.[2] Currently, 21 food product prices are still regulated. Examples include white (soft) cheese, several types of bread, etc. The government caps the regulated products' prices. Retailers are not allowed to sell them at a price exceeding the cap price.[3] Importantly, the cap prices last long periods. In addition, retailers tend to sell these products at cap prices (even though they could sell them

---

[1] See, for example, Mankiw and Reis (2002, 2010), Zbaracki et al. (2004), Ball et al. (2005), Reis (2006a, 2006b) and (2009), Keen (2007), Klenow and Willis (2007), Falkinger (2008), Knotek (2010), Konieczny (2007), and Dhyne and Konieczny (2014).
[2] See, for example, Lach and Tsiddon (1992, 1996, and 2007), Sargent and Zeira (2011), Avishay-Rizi and Ater (2021), Ater and Avishay-Rizi (2022), and Ater and Gerlitz (2017).
[3] For more details about price controls in Israel, their rationale, and their history, see Hagai (2009).

at lower prices), implying that there is little variation in their prices.

## 3. Data

*3.1. Data description*

We use three datasets. Our main dataset comes from consumer surveys that we ran in two rounds, one in the first quarter of 2016, and the second in the first and second quarters of 2019. Shoppers exiting stores were given a list of products and asked to identify the products they had purchased on their current shopping trip. They were also asked to recall the prices of the products. Additionally, they were asked to provide general socio-demographic information and details about their shopping habits.[4] Finally, they were given a list of products and asked to indicate whether their prices were regulated or not. The other two datasets are used in robustness analysis and are described in Appendix C.

*3.2. Survey design*

To adapt the shopper survey to a natural experimental design, it was necessary to collect data on both regulated and comparable unregulated products. We started by considering the 21 products whose prices are regulated. Some of these products are similar, however. For example, there are five types of regulated bread (white, whole wheat, sliced white, sliced whole wheat, and challah). To shorten the questionnaire, we excluded such duplicate products. We ended up selecting 10 products whose prices are regulated.

For each regulated product, we chose a comparable product, i.e., products with similar properties, but whose prices are not regulated. For example, White (soft) Spread Cheese, which is subject to price control, is not a perfect substitute for Cottage Cheese, which is not subject to price control, but they come in the same size (250 gr.) and are usually consumed in the same form. Indeed, both are staple food items, and most Israeli households purchase them regularly (Hendel et al., 2017). As another example, butter, which is subject to price control, and margarine, which is not, are quite comparable in their size, in their use, etc. In addition, both are staple items. Table 1 lists the 10 matched pairs of sampled products.

*3.3. Descriptive statistics*

The pooled sample consists of 855 shoppers surveyed at 13 different supermarkets. Table 2 presents summary statistics of the shoppers' socio-demographic profiles, shopping habits, and their accuracy in recalling prices, for the surveys in rounds 1 and 2, and the pooled sample. The price recall accuracy is measured as the average absolute value of the percentage error, $|\ln(recall\ price_{i,j,t}/actual\ price_{j,t})|$, where $recall\ price_{i,j,t}$ is the price recalled by participant $i$, of product $j$ surveyed at time $t$, and $actual\ price_{j,t}$ is the posted price of product $j$ at time $t$.

---
[4] A translation of the questionnaire is presented in Appendix G.

When asked to classify which products are subject to price controls, participants were correct in about 2/3 of the cases, which is statistically greater than the 50% predicted under random responses ($t = 26.64$, $p < 0.01$).

The average absolute value of the price recall error is 27.3%. Comparing regulated and unregulated products, we find that the average absolute value of the price recall errors of regulated and unregulated products are 30.8% and 22.3%, respectively. The difference is statistically significant at the 10% level ($t = 1.68$).[5]

## 4 Estimation results

Below we present the results for the pooled data. In Appendix A, we show that the results hold also if we analyze the data of the two rounds separately.

To test the participants' attention to prices, we treat the errors the consumers make in recalling prices as a proxy for the attention they pay to price information. We hypothesize that consumers make larger errors in recalling product prices if they pay less attention to them. According to the sticky information model, people will pay less attention to prices that change infrequently, i.e., to prices of regulated products. The model, therefore, predicts that participants will make larger errors in recalling the prices of regulated products in comparison to comparable unregulated products.

Table 3 reports the results of a set of regressions testing this hypothesis. We report robust standard errors, clustered at the participants' level. The dependent variable in all regressions is the absolute value of the percentage error the shoppers make in recalling the prices. In column 1, the only covariate is a dummy for price control (1 = subject to price control, 0 = otherwise), in addition to supermarket and product-pair fixed effects. We also include random effects for participants. The estimated coefficient of the price control dummy is statistically significant at the 1% level, and its magnitude suggests that shoppers make errors that are 3.4% larger in recalling the prices of regulated products than when recalling the prices of unregulated products. This amounts to about 12.5% of the average absolute value of the price recall error of 27.3%.

In column 2, we add controls for the products' actual posted prices, the shopper's age, the household size, the number of cars the household possesses, the number of supermarkets the shoppers patronize, the average amount of money spent by the shopper on a shopping trip, and the proportion of the sampled 20 products which the shoppers correctly identified as regulated and unregulated. In addition, the regression includes dummy variables for gender (1 = woman, 0 = otherwise), marital status (1 = married, 0 = otherwise), academic degree (1 = with academic degree, 0 = otherwise), ultra-religious (1 = the shopper identifies himself/herself as ultra-religious, 0 = otherwise), frequent buyer (1 = shops more than once a week, 0 = otherwise), and for products with .90-ending prices.

---
[5] In the data, there are some observations with unreasonably large price recall errors. Therefore, to be conservative, we have decided to exclude 103 observations (about 3.2% of the total), with recall errors greater than 100%. In Appendix B, we show that our results are robust to changing the threshold.

The estimated coefficient of the price control dummy, 4.1%, is larger than in the previous estimation, and it remains statistically significant at the 1% level.

In column 3, we test the robustness of these results by adding fixed effects for the interactions between the supermarkets and the product pairs, which capture systematic differences in the attention that shoppers who patronize different supermarkets pay to each product. We find that the coefficient of the regulated products remains at 4.1%.

A possible explanation for our findings is that shoppers who buy regulated products do not buy unregulated ones. In that case, the differences we find might be due to differences between shoppers rather than differences between the decisions that shoppers make regarding attention to the price information. In column 4, therefore, we control for this by estimating a fixed effects regression controlling for the interaction between consumers and product categories. The identification in this regression is based on shoppers who bought both regulated and unregulated products belonging to the same pairs (e.g., white cheese and cottage cheese). Using fixed effects regression leads to dropping all the coefficients that do not vary within participants. We find that the effect of the regulated prices remains positive ($\beta = 0.063$) and statistically significant.

## 5. Robustness tests, caveats, and conclusion

Our results support a key prediction of the sticky information model: shoppers make larger errors when recalling prices of regulated products than comparable unregulated products. In Appendix C, we show that these findings cannot be explained by products' characteristics: (1) the prices of regulated products last longer than the prices of comparable unregulated products, (2) the price dispersion of the regulated products is smaller than the price dispersion of the comparable unregulated products, and (3) the shoppers purchase the regulated products at least as often, and usually more often, than the unregulated products.

We might conjecture that it should be easier for the shoppers to recall prices that (i) remain unchanged for long periods, and (ii) they encounter on a regular basis. Then, if shoppers were to give the same amount of attention to all products, then they should have recalled the prices of regulated products with greater accuracy than the prices of comparable unregulated products, which is counter to what we find.

Importantly, our results hold when we control for possible alternative explanations. First, it is possible that shoppers are more attentive to unregulated product prices because they are more expensive than comparable regulated products. However, the inclusion of the products' prices as a control variable does not change our findings. Further, in Appendix D, we show that focusing on pairs of products with comparable prices does not alter the results.

Second, it is possible that because we measure the price recall error in relative terms, the size of the errors appears larger because the regulated products tend to have lower prices than comparable unregulated products. To account for this possibility, we show in Appendix E that the results remain unchanged when we use levels rather than relative price recall errors.

Third, our results might be affected by the special attention that shoppers, and the media, give to cottage cheese (Hendel et al., 2017). In Appendix F, we show, however, that excluding observations on the cottage cheese – soft cheese pair does not alter the results.

Fourth, shoppers who buy regulated products may have different characteristics than those who buy unregulated products. However, adding fixed effects for the interactions between product pairs and shoppers in column 4 of Table 3 doesn't alter the results.

Therefore, the finding that consumers make larger errors in recalling the prices of regulated products is an indication that consumers are less attentive to information that changes infrequently, and with small dispersion, as argued by Reis (2006b).

Alternatively, as the reviewer has noted, shoppers may pay less attention to regulated products because they have different beliefs about regulated vs. unregulated products. For example, they might believe that the regulators guarantee "a low price." We cannot rule out such explanations, but we show in Appendix D that the results hold for pairs of products with similar prices, and even in cases where the regulated product prices are higher than the prices of comparable unregulated products. Our findings may also be consistent with certain models of rational inattention, although these models often share key components with the sticky information model.

More work is necessary to determine the model that best fits the data. However, the main empirical finding we report is robust – shoppers seem to pay differential attention to price information. Further, the level of attention that they pay varies with the expected benefit from processing the information – as predicted by models of sticky information.

The inattention that shoppers exhibit towards the prices of regulated products is likely costly because it likely leads to suboptimal consumption decisions. However, it is unclear whether making more frequent price changes that will incentivize consumers to pay greater attention to the prices will necessarily increase consumer welfare because they will have to incur higher costs for being more attentive (Reis, 2006b).

Overall, our findings are consistent with the argument that consumers make rational choices about how much information to collect. Furthermore, the results have implications for assessing the effect of the price control regulation: the regulation's goal is to maintain low prices and thus enhance consumers' welfare. However, if consumers possess less information about the prices of regulated products compared to comparable unregulated products, then some of the intended benefits may be lost as consumers fail to optimize their consumption basket.

**Acknowledgments**

We are grateful to the anonymous reviewer for helpful and constructive comments. This is a substantially revised version of the manuscript we presented at the Israeli Economic Association Annual Conference. We thank the participants of the conference and especially Alon Eizenberg, for their helpful discussions and suggestions. The paper is based on Chapter 3 of Doron Sayag's PhD dissertation at Bar-Ilan University. All errors are ours.

# References

Ater, I., Avishay-Rizi, O., 2022. Price saliency and fairness: Evidence from regulatory shaming. CEPR Discussion Paper No. DP17156.

Avishay-Rizi, O., Ater, I., 2021. Fairness considerations in pricing decisions: Evidence from shaming regulation. Manuscript, presented at the 2021 IEA annual Conference.

Ater, I., Gerlitz, O., 2017. Round prices and price rigidity: Evidence from outlawing odd prices. Journal of Economic Behavior and Organization 144(C), 188–203.

Ball, L., Mankiw, N.G., Reis, R., 2005. Monetary policy for inattentive economies. Journal of Monetary Economics 52(4), 703–725.

Dexter, A., Levi, M., Nault, B., 1993. Freely determined versus regulated prices: Implications for the measured link between money and inflation. Journal of Money, Credit and Banking 25(2), 222–230.

Dexter, A., Levi, M., Nault, B., 2002. Sticky prices: the impact of regulation. Journal of Monetary Economics 49, 797–821.

Dhyne, E., Konieczny, J., 2014. Aggregation and the staggering of price changes. Economic Inquiry 52(2), 732–756.

Dickson, P.R., Sawyer, A.G., 1990. The price knowledge and search of supermarket shoppers. Journal of Marketing 54(3), 42–53.

Falkinger, J., 2008. Limited attention as a scarce resource in information-rich economies. Economic Journal 118(532), 1596–1620.

Hagai, Z., 2009. Goods and services under price control: A survey and analysis of their impact on households. Division of Econ. and Research, the Ministry of Industry, Trade, and Employment, http://economy.gov.il/Research/Documents/X9121.pdf (in Hebrew).

Hendel, I., Lach, S., Spiegel, Y., 2017). Consumers' activism: The cottage cheese boycott. Rand Journal of Economics 48(4), 972–1003.

Keen, B., 2007. Sticky price and sticky information price-setting models: What is the difference? Economic Inquiry 45(4), 770–786.

Klenow, P., Willis, J., 2007. Sticky information and sticky prices. Journal of Monetary Economics 54 (Supplement), 79–99.

Knotek, E., II, 2010. A tale of two rigidities: Sticky prices in a sticky information environment. Journal of Money, Credit, and Banking 42(8), 1543–1564.

Konieczny, J., 2007. Costly price adjustment and the optimal rate of inflation. Managerial and Decision Economics 28(6), 591–603.

Lach, S., Tsiddon, D., 1992. The behavior of prices and inflation: An empirical analysis of disaggregated data. Journal of Political Economy 100(2), 349–389.

Lach, S., Tsiddon, D., 1996. Staggering and synchronization in price-setting: Evidence from multiproduct firms. American Economic Review 86, 1175–1196.

Lach, S., Tsiddon, D., 2007. Small price changes and menu costs. Managerial and Decision Economics 28, 649–656.

Mankiw, N.G., Reis, R., 2002. Sticky Information versus sticky prices: A proposal to replace the New Keynesian Phillips Curve. Quarterly Journal of Economics 117(4), 1295–1328.

Mankiw, N. G., Reis, R., 2010. Imperfect information and aggregate supply, in B. Friedman and M. Woodford (eds.), Handbook of Monetary Economics, Volume 3 (New York, NY: Elsevier), pp. 183–229.

Reis, R., 2006a. Inattentive producers. Review of Economic Studies 73(3), 793–821.

Reis, R., 2006b. Inattentive consumers. Journal of Monetary Economics 53(8), 1761–1800.

Reis, R., 2009. Optimal monetary policy rules in an estimated sticky-information model. American Economic Journal: Macroeconomics 1(2), 1–28.

Sargent, T.J., Zeira, Y., 2011. Israel 1983: A bout of unpleasant monetarist arithmetic. Review of Economic Dynamics 14, 419–431.

Urbany, J.E., Dickson, P.R., Kalapurakal, R., 1996. Price search in the retail grocery market. Journal of Marketing 91–104.

Zbaracki, M.J., Ritson, M., Levy, D., Dutta, S., Bergen, M., 2004. Managerial and customer costs of price adjustment: Direct evidence from industrial markets. Review of Economics and Statistics 86(2), 514–553.

**Table 1**

Matched pairs of comparable products: Products subject to price control regulation (LHS) and products not subject to price control regulation (RHS).

| Category No. | Products subject to price control regulation | Products not subject to price control regulation |
|---|---|---|
| 1. | Soft cheese ("white cheese') | Cottage cheese |
| 2. | Semi-hard cheese - sold by weight | Prepackaged semi-hard cheese |
| 3. | Soured milk | Plain yoghurt |
| 4. | Butter | Margarine |
| 5. | Whip cream 38% | Non-dairy whip cream |
| 6. | Packaged milk (3%) | Soy milk |
| 7. | Eggs (large) | Free/ omega eggs |
| 8. | White bread – sliced | Pita bread |
| 9. | Whole wheat bread – sliced | Light bread – sliced |
| 10. | Salt, 1 kg | Sugar / Black pepper |

Notes

In the first round of the survey, we used sugar as a "comparable" product to salt. In the second round, we replaced sugar with black pepper.

**Table 2**

Summary statistics of the shoppers sampled.

| Variable | Round 1 | Round 2 | Pooled sample |
| --- | --- | --- | --- |
| Average age | 37.9 (15.6) | 36.1 (13.2) | 37.1 (14.6) |
| Average household size | 3.3 (1.97) | 2.9 (1.75) | 3.1 (1.89) |
| Female | 57.6% | 56.1% | 58.4% |
| Academic | 61.8% | 42.8% | 53.9% |
| Married | 51.2% | 60.2% | 51.9% |
| Ultra-religious | 3.2% | 7.3% | 4.9% |
| Average no. of cars owned | 1.2 (0.90) | 1.2 (0.86) | 1.2 (0.87) |
| Average amount spent per shopping trip (in NIS) | 363.5 (216.0) | 366.1 (210.2) | 365.7 (216.3) |
| Average no. of shops visited | 2.16 (0.8) | 1.78 (0.74) | 01.2 (0.87) |
| Shopping more than once a week | 23.6% | 20.8% | 22.5% |
| Correctly identifying whether the products sampled are subject to price regulation or not | 65.4% | 69.8% | 67.3% |
| Average absolute value of the recall error | 29.7% | 24.2% | 27.3% |
| Average absolute value of the recall error, products subject to price control | 35.3% | 26.0% | 30.8% |
| Average absolute value of the recall error, products *not* subject to price control | 23.1% | 20.8% | 22.3% |
| No. of products purchased per shopper | 3.6 (1.94) | 4.1 (2.95) | 3.8 (2.42) |
| No. of food retailers sampled | 5 | 8 | 13 |
| No. of shoppers | 500 | 355 | 855 |
| No. of observations | 1,819 | 1,446 | 3,265 |

Notes

Round 1 of the survey was conducted in the first quarter of 2016. Round 2 of the survey was conducted in the first two quarters of 2019.

**Table 3**

Estimation results: pooled sample.

| Variable | (1) | (2) | (3) | (4) |
|---|---|---|---|---|
| Price control (dummy) | 0.034*** | 0.041*** | 0.041*** | 0.063*** |
|  | (0.009) | (0.010) | (0.010) | (0.020) |
| Product's posted price |  | 0.002 | 0.003** | 0.001 |
|  |  | (0.001) | (0.001) | (0.004) |
| Age |  | −0.001*** | −0.001*** |  |
|  |  | (0.000) | (0.000) |  |
| Household size |  | −0.001 | −0.001 |  |
|  |  | (0.002) | (0.002) |  |
| No. of cars |  | 0.005 | 0.007 |  |
|  |  | (0.006) | (0.006) |  |
| No. of supermarkets visited |  | −0.005 | −0.006 |  |
|  |  | (0.005) | (0.005) |  |
| Average amount spent |  | 0.001 | −0.001 |  |
|  |  | (0.003) | (0.003) |  |
| Woman (dummy) |  | −0.006 | −0.005 |  |
|  |  | (0.008) | (0.008) |  |
| Married (dummy) |  | −0.011 | −0.007 |  |
|  |  | (0.008) | (0.008) |  |
| Academic degree (dummy) |  | −0.006 | −0.008 |  |
|  |  | (0.008) | (0.008) |  |
| Ultra-religious (dummy) |  | 0.001 | 0.012 |  |
|  |  | (0.020) | (0.020) |  |
| Frequent shopper (dummy) |  | −0.005 | −0.008 |  |
|  |  | (0.010) | (0.010) |  |
| Percentage of correct recalls of regulated prices |  | −0.088* | −0.056 |  |
|  |  | (0.047) | (0.047) |  |
| .90-ending price (dummy) |  | 0.002 | −0.013 | −0.061** |
|  |  | (0.010) | (0.013) | (0.029) |
| Constant | 0.195*** | 0.273*** | 0.214*** | −0.64* |
|  | (0.021) | (0.039) | (0.047) | (0.364) |
| $R^2$ | 0.24 | 0.25 | 0.35 | 0.01 |
| $N$ | 3,162 | 3,162 | 3,162 | 3,162 |

Notes

1. In all columns, the dependent variable is $ln \left| \frac{Recall.price_{i,j,t}}{Actual.price_{j,t}} \right|$, where $i$ stands for the participant, $j$ for a product, and $t$ for the week of the observation.
2. Columns 1–3 report the results of regressions with random effects for participants.
3. Column 4 reports the regression results with fixed effects for participants × product pairs.
4. The standard errors are robust and clustered at the participants' level.
5. All the regressions include product-pair and supermarket dummies (not reported to save space).
6. Observations with relative errors above 100% were excluded from the regressions.
7. ***, **, and * indicate significance at 1%, 5%, and 10%, respectively.

**Online Supplementary Web Appendix**

# Sticky Information and Price Controls: Evidence from a Natural Experiment

Doron Sayag
Department of Economics, Bar-Ilan University
Ramat-Gan 5290002, Israel
Doronsayag2@gmail.com

Avichai Snir
Department of Economics, Bar-Ilan University
Ramat-Gan 5290002, Israel
Snirav@biu.ac.il

Daniel Levy
Department of Economics, Bar-Ilan University
Ramat-Gan 5290002, Israel,
Department of Economics, Emory University
Atlanta, GA 30322, USA,
ICEA, ISET at TSU, and RCEA
Daniel.Levy@biu.ac.il

Revised:
February 18, 2025

0

**Appendix A. Estimation results for each round separately**

We had two rounds of the survey. The first was conducted in the 1st quarter of 2016, and the second in the 1st and 2nd quarters of 2019. In the paper, we present results using pooled data. In this section, we estimate separate regressions for each round.

Table A1 summarizes the results of round 1 and Table A2—the results of round 2. We report robust standard errors, clustered at the participants' level. The dependent variable in all regressions is the shoppers' absolute value of the percentage error in recalling the prices. In column 1, the only covariate is a dummy for price control (1 = subject to price control, 0 = otherwise), in addition to supermarket and product-pair fixed effects. We also include random effects for participants.

In Table A1 (A2) the estimated coefficient of the price control dummy is statistically significant at the 1% (5%) level, and its magnitude suggests that shoppers make errors that are 3.7% (2.5%) larger in recalling the prices of regulated products than when recalling the prices of unregulated products. This amounts to about 12.5% (10.3%) of the average absolute value of the price recall error of 29.7% (24.2%).

In column 2, we add controls for the products' actual posted prices, the shopper's age, the household size, the number of cars the household possesses, the number of supermarkets the shoppers patronize, the average amount of money spent on a shopping trip, and the proportion of the sampled 20 products which the shoppers correctly identified as regulated/unregulated. In addition, the regression includes dummy variables for gender (1 = woman, 0 = otherwise), marital status (1 = married, 0 = otherwise), academic degree (1 = with academic degree, 0 = otherwise), ultra-religious (1 = the shopper identifies himself/herself as ultra-religious, 0 = otherwise), frequent buyer (1 = shops more than once a week, 0 = otherwise), and for products with .90-ending prices.

The estimated coefficient in Table A1 (A2) of the price control dummy, 5.4% (1.6%), is statistically significant at the 1% level (is not statistically significant).

In column 3, we test the robustness of these results by adding fixed effects for the interactions between the supermarkets and the product pairs. These dummies capture systematic differences in the attention that shoppers who patronize different supermarkets pay to each product. We find that in Table A1 (A2) the coefficient of the regulated products is 6.3% (1.0%) and is statistically significant at the 1% (5%) level.

A possible explanation for our findings is that shoppers who buy regulated products do not buy unregulated ones. In that case, the differences might be due to differences between shoppers rather than differences in their attention to the price information. In column 4, we account for this by estimating fixed effects regression controlling for the interaction between consumers and product categories. The identification in this regression is based on shoppers who bought both regulated and unregulated products belonging to the same pairs (e.g., white cheese and cottage cheese). Using fixed effects regression leads to dropping all the coefficients that do not vary within participants. We find that in Table A1 (A2) the effect of the regulated prices is 5.6% (2.7%) and statistically significant at the 1% (5%) level.

Table A1. Estimation results – round 1

| Variable | (1) | (2) | (3) | (4) |
|---|---|---|---|---|
| Price control (dummy) | 0.037** (0.012) | 0.054*** (0.013) | 0.063*** (0.014) | 0.056*** (0.021) |
| Product's posted price |  | 0.004* (0.002) | 0.005** (0.002) | −0.003 (0.004) |
| Age |  | −0.001** (0.000) | −0.001* (0.000) |  |
| Household size |  | 0.002 (0.003) | 0.003 (0.003) |  |
| No. of cars |  | 0.008 (0.009) | 0.008 (0.008) |  |
| No. of supermarkets visited |  | −0.004 (0.006) | −0.004 (0.006) |  |
| Average amount spent |  | −0.045 (0.065) | −0.005 (0.004) |  |
| Woman (dummy) |  | −0.006 (0.011) | −0.005 (0.011) |  |
| Married (dummy) |  | −0.011 (0.012) | −0.010 (0.012) |  |
| Academic degree (dummy) |  | 0.004 (0.012) | −0.001 (0.012) |  |
| Ultra-religious (dummy) |  | 0.030 (0.039) | 0.029 (0.037) |  |
| Frequent shopper (dummy) |  | −0.018 (0.013) | −0.019 (0.013) |  |
| Percentage of correct recalls of regulated prices |  | −0.045 (0.065) | −0.047 (0.065) |  |
| .90-ending price (dummy) |  | −0.012 (0.013) | −0.023 (0.017) | −0.078** (0.032) |
| Constant | 0.174*** (0.022) | 0.205*** (0.050) | 0.186*** (0.056) | 0.208*** (0.033) |
| $R^2$ | 0.22 | 0.23 | 0.28 | 0.005 |
| N | 1,775 | 1,775 | 1,775 | 1,775 |

Notes:

1. In all columns, the dependent variable is $ln\left|\frac{Recall.price_{i,j,t}}{Actual.price_{j,t}}\right|$, where $i$ stands for the participant, $j$ for a product, and $t$ for the week of the observation.
2. Columns 1–3 report the results of regressions with random effects for participants. Column 4 reports the regression results with fixed effects for participants×product pairs.
4. The standard errors are robust and clustered at the participants' level.
5. All the regressions include product-pair and supermarket dummies (not reported to save space).
6. Observations with relative errors above 100% were excluded from the regressions.
7. Using data from round 1 of the survey.
7. ***, **, and * indicate significance at 1%, 5%, and 10%, respectively.

Table A2. Estimation results – round 2

| Variable | (1) | (2) | (3) | (4) |
|---|---|---|---|---|
| Price control (dummy) | 0.025** (0.012) | 0.016 (0.014) | 0.010** (0.000) | 0.027** (0.012) |
| Product's posted price |  | −0.003 (0.002) | −0.000 (0.001) | 0.028** (0.011) |
| Age |  | −0.001*** (0.001) | −0.001** (0.000) |  |
| Household size |  | −0.009*** (0.003) | −0.009*** (0.003) |  |
| No. of cars |  | 0.007 (0.008) | 0.010 (0.007) |  |
| No. of supermarkets visited |  | −0.006 (0.007) | −0.006 (0.007) |  |
| Average amount spent |  | 0.010* (0.006) | 0.008 (0.005) |  |
| Woman (dummy) |  | −0.007 (0.012) | -0.005 (0.011) |  |
| Married (dummy) |  | −0.007 (0.012) | 0.002 (0.011) |  |
| Academic degree (dummy) |  | −0.017 (0.010) | −0.017* (0.010) |  |
| Ultra-religious (dummy) |  | −0.022 (0.021) | -0.015 (0.020) |  |
| Frequent shopper (dummy) |  | 0.013 (0.015) | 0.011 (0.015) |  |
| Percentage of correct recalls of regulated prices |  | −0.162** (0.075) | −0.126* (0.071) |  |
| .90-ending price (dummy) |  | 0.048** (0.018) | 0.013 (0.020) | −0.119 (0.075) |
| Constant | 0.019 (0.016) | 0.048*** (0.018) | 0.179*** (0.050) | −0.049 (0.087) |
| $R^2$ | 0.29 | 0.31 | 0.45 | 0.01 |
| N | 1,387 | 1,387 | 1,387 | 1,387 |

Notes:

1. In all columns, the dependent variable is $ln\left|\frac{Recall.price_{i,j,t}}{Actual.price_{j,t}}\right|$, where $i$ stands for the participant, $j$ for a product, and $t$ for the week of the observation.
2. Columns 1–3 report the results of regressions with random effects for participants. Column 4 reports the regression results with fixed effects for participants×product pairs.
4. The standard errors are robust and clustered at the participants' level.
5. All the regressions include product-pair and supermarket dummies (not reported to save space).
6. Observations with relative errors above 100% were excluded from the regressions.
7. Using data from round 2 of the survey.
7. ***, **, and * indicate significance at 1%, 5%, and 10%, respectively.

**Appendix B. Changing the threshold of the maximum absolute value of the price recall error**

In the paper, we restrict the absolute value of the percentage errors that participants made in recalling the prices of products to less than 100%. Below, we show that the results are robust to changing this threshold. First, Table B1 reports the results of a series of regressions that we estimate using absolute value of the percentage errors smaller than 50%. Setting this restriction forces us to drop 437 observations (13.8%).

We report robust standard errors, clustered at the participants' level. The dependent variable in all regressions is the absolute value of the percentage error the shoppers made in recalling the prices. In column 1, the only covariate is a dummy for price control (1 = subject to price control, 0 = otherwise), in addition to supermarket and product-pair fixed effects. We also include random effects for participants. In Table B1, the estimated coefficient of the price control dummy is statistically significant at the 1% level, and its magnitude suggests that shoppers make errors that are 1.8% larger in recalling the prices of regulated products than when recalling the prices of unregulated products. This amounts to about 14.5% of the average absolute value of the price recall error, conditional on the average absolute value of the price recall error < 50%, of 12.4%.

In column 2, we add controls for the products' actual posted prices, the shopper's age, the household size, the number of cars the household possesses, the number of supermarkets the shoppers patronize, the average amount of money spent on a shopping trip, and the proportion of the sampled 20 products which the shoppers correctly identified as regulated and unregulated. In addition, the regression includes dummy variables for gender (1 = woman, 0 = otherwise), marital status (1 = married, 0 = otherwise), academic degree (1 = with academic degree, 0 = otherwise), ultra-religious (1 = the shopper identifies himself/herself as ultra-religious, 0 = otherwise), frequent buyer (1 = shops more than once a week, 0 = otherwise), and for products with .90-ending prices.

The estimated coefficient of the price control dummy, 1.6%, is statistically significant at the 1% level. In column 3, we test the robustness of these results by adding fixed effects for the interactions between the supermarkets and the product pairs. These dummies capture systematic differences in the attention that shoppers who patronize different supermarkets pay to each product. We find that the coefficient of the regulated products is 2.1% and is statistically significant at the 1% level.

A possible explanation for our findings is that shoppers who buy regulated products do not buy unregulated ones. In that case, the differences we find might be due to differences between shoppers rather than differences in their attention to the price information. In column 4, we control for this by estimating a fixed effects regression controlling for the interaction between consumers and product categories. The identification in this regression is based on shoppers who bought both regulated and unregulated products belonging to the same pairs (e.g., white cheese and cottage cheese). Using fixed effects regression leads to dropping all the coefficients that do not vary within participants. We find that the effect of the regulated prices is 3.0% and statistically significant at the 5% level.

As another test, we estimated the same regressions but without excluding any observation. The results are summarized in Table B2. In column 1, the coefficient of the regulated products dummy is 0.063. In column 2, it is 0.059. In column 3, It is 0.059. In column 4, it is 0.101. In all columns, these coefficients are statistically significant at 1%.

We, therefore, conclude that using a stricter rule for excluding observations or choosing to include all observations does not change our main finding. Shoppers tend to make larger errors in recalling the prices of regulated products.

Table B1. Estimation results – Absolute value of the relative recall errors smaller than 50%

| Variable | (1) | (2) | (3) | (4) |
|---|---|---|---|---|
| Price control (dummy) | 0.018*** | 0.016*** | 0.021*** | 0.030** |
| | (0.005) | (0.006) | (0.006) | (0.013) |
| Product's posted price | | −0.000 | 0.001 | 0.004* |
| | | (0.001) | (0.001) | (0.002) |
| Age | | −0.000*** | −0.000*** | |
| | | (0.000) | (0.000) | |
| Household size | | −0.001 | −0.001 | |
| | | (0.001) | (0.001) | |
| No. of cars | | 0.001 | 0.002 | |
| | | (0.004) | (0.003) | |
| No. of supermarkets visited | | −0.002 | −0.002 | |
| | | (0.003) | (0.003) | |
| Average amount spent | | 0.002 | 0.001 | |
| | | (0.002) | (0.002) | |
| Woman (dummy) | | −0.001 | 0.001 | |
| | | (0.005) | (0.005) | |
| Married (dummy) | | −0.006 | −0.007 | |
| | | (0.005) | (0.005) | |
| Academic degree (dummy) | | −0.007 | −0.008 | |
| | | (0.005) | (0.005) | |
| Ultra-religious (dummy) | | 0.015 | 0.016 | |
| | | (0.013) | (0.012) | |
| Frequent shopper (dummy) | | 0.002 | 0.002 | |
| | | (0.006) | (0.006) | |
| Percentage of correct recalls of regulated prices | | −0.064** | −0.048* | |
| | | (0.029) | (0.029) | |
| .90-ending price (dummy) | | 0.004 | 0.008 | −0.032* |
| | | (0.006) | (0.008) | (0.018) |
| Constant | 0.150*** | 0.209*** | 0.159*** | −0.066 |
| | (0.012) | (0.026) | (0.031) | (0.240) |
| $R^2$ | 0.19 | 0.20 | 0.28 | 0.004 |
| $N$ | 2,828 | 2,828 | 2,828 | 2,828 |

Notes:

1. In all columns, the dependent variable is $ln\left|\frac{Recall.price_{i,j,t}}{Actual.price_{j,t}}\right|$, where $i$ stands for the participant, $j$ for a product, and $t$ for the week of the observation.
2. Columns 1–3 report the results of regressions with random effects for participants. Column 4 reports the regression results with fixed effects for participants×product pairs.
4. The standard errors are robust and clustered at the participants' level.
5. All the regressions include product-pair and supermarket dummies (not reported to save space).
6. Observations with absolute values of the relative errors above 50% were excluded from the regressions.
7. Using data from the pooled sample.
7. ***, **, and * indicate significance at 1%, 5%, and 10%, respectively.

Table B2. Estimation results – All observations

| Variable | (1) | (2) | (3) | (4) |
|---|---|---|---|---|
| Price control (dummy) | 0.063*** (0.012) | 0.059*** (0.014) | 0.059*** (0.013) | 0.101*** (0.028) |
| Product's posted price | | −0.001 (0.002) | 0.004** (0.002) | 0.003 (0.004) |
| Age | | −0.001*** (0.000) | −0.001** (0.000) | |
| Household size | | −0.002 (0.003) | −0.002 (0.003) | |
| No. of cars | | 0.012 (0.009) | 0.015* (0.010) | |
| No. of supermarkets visited | | −0.010 (0.007) | −0.013 (0.007) | |
| Average amount spent | | 0.002 (0.004) | −0.001 (0.004) | |
| Woman (dummy) | | −0.010 (0.010) | −0.011 (0.10) | |
| Married (dummy) | | −0.014 (0.011) | −0.010 (0.011) | |
| Academic degree (dummy) | | 0.005 (0.011) | 0.002 (0.011) | |
| Ultra-religious (dummy) | | 0.016 (0.025) | 0.015 (0.025) | |
| Frequent shopper (dummy) | | −0.007 (0.013) | -0.008 (0.013) | |
| Percentage of correct recalls of regulated prices | | −0.173*** (0.007) | −0.145** (0.066) | |
| .90-ending price (dummy) | | 0.019 (0.015) | −0.011 (0.017) | −0.077* (0.041) |
| Constant | 0.255*** (0.036) | 0.412*** (0.061) | 0.322*** (0.069) | −1.277** (0.458) |
| $R^2$ | 0.27 | 0.27 | 0.39 | 0.001 |
| $N$ | 3,265 | 3,265 | 3,265 | 3,265 |

Notes:

1. In all columns, the dependent variable is $ln\left|\frac{Recall.price_{i,j,t}}{Actual.price_{j,t}}\right|$, where $i$ stands for the participant, $j$ for a product, and $t$ for the week of the observation.
2. Columns 1–3 report the results of regressions with random effects for participants. Column 4 reports the regression results with fixed effects for participants×product pairs.
4. The standard errors are robust and clustered at the participants' level.
5. All the regressions include product-pair and supermarket dummies (not reported to save space).
6. Including all observations.
7. Using data from the pooled sample.
7. ***, **, and * indicate significance at 1%, 5%, and 10%, respectively.

**Appendix C. The characteristics of regulated and unregulated products**

A possible concern is that shoppers are less precise in recalling the prices of products subject to price controls because they are less familiar with these products, rather than because they pay less attention to their prices. This might happen if (1) prices of regulated products change often, making it hard for shoppers to recall the current price, (2) if there is a large cross-store variation in the price of products whose prices are subject to control, or (3) if the shoppers rarely buy products whose prices are capped.

To rule out these possibilities, we run several tests using two other datasets, the CPI dataset, and A.C. Nielsen's datasets.

**Data**

The CPI dataset includes price data collected during 2016–2021 by Israel's Central Bureau of Statistics (CBS) for compiling the CPI. The data includes information on the prices of individual products offered in a representative sample of stores. The data is collected monthly by surveyors who visit the stores and collect the price information.

The A.C. Nielsen dataset is a retail scanner data for 2018, which contains monthly observations on revenues and sales volumes of products in Fast-Moving Consumer Goods (FMCG) markets. The data, which are obtained directly from individual retailer scanners, are aggregated to produce national-level data. Thus, an observation might indicate that a specific brand of orange juice sold 19,543 units during a particular month, generating a total revenue of NIS 140,123. Nielsen data covers about 96% of the FMCG market.

**Results**

First, we use the CPI data to compute the average duration between price changes for each product, calculated as $-N_j^{-1} \sum \left[ ln(1 - \bar{f}_{js}) \right]^{-1}$ (Nakamura and Steinsson, 2008). In this setting, $\bar{f}_{js}$ is the average frequency of price changes for product $j$ in store $s$.

Table C1 gives the results. The LHS panel gives the average duration, in months, between price changes of products whose prices are subject to controls. The RHS panel gives the corresponding statistics of comparable products whose prices are not subject to controls. Except for the category of "white bread–sliced," which has a duration that is similar to that of pita bread, in all other cases the prices of products whose prices are regulated, last much longer than the prices of the comparable but unregulated products.

Second, using the CPI data we look at the variance of the prices of products whose prices are subject to controls. Figure C1 shows that in over 70% of cases, the products are offered at cap prices. We also find that most of the deviations from cap prices are small, up to 3% in absolute values. These small deviations are found mostly in stores that round the cap prices when the latter are not 0-ending.[1]

This suggests that the variance of the prices of products subject to price control is low. Table C2 gives summary statistics of the distributions of the prices of products whose prices are subject to controls (LHS panel) and of prices of products whose prices are not subject to controls (RHS panel). For all pairs, the standard deviations and the coefficients of variation (CV) of the products not subject to price controls are much larger than the corresponding values of the products subject to price controls.[2]

Finally, we consider whether shoppers might pay less attention to the prices of products subject to price controls because these products are not popular. We use Nielsen data to calculate, for each pair of products, the total number of units bought in 2018. We then find the share of the units of products that are subject to price controls out of the total.

Figure C2 depicts the results. We find that for 9 of the 10 pairs, the share of the products whose prices are subject to control is at least 35%. For 6 of these 9 pairs, the share of the products whose prices are subject to control is at least 60%.

We, therefore, conclude that the differences between the ability of shoppers to recall the prices of products whose prices are and aren't subject to control cannot be explained by greater difficulty in recalling these prices, or by shoppers viewing the products as unimportant. On the contrary, these products are usually more popular than comparable products whose prices are not subject to controls, their prices remain stable for longer periods, and there is less variation in their prices across stores.

---

[1] Since January 2014, the prices of all packaged products sold in Israel must be 0-ending (Ater and Gerlitz 2017, Sayag et al. 2024, Snir et al. 2017). The only exceptions are products whose prices are subject to control because the cap prices that the government sets are not necessarily round. It turns out that some stores prefer to round the prices to the nearest 0-ending price. Occasionally, they round the prices upwards, which is illegal.

[2] The variation and the coefficients of variation for products whose prices are subject to controls are low also because such products usually come in a single package size. Similar products that are not subject to price controls often come in different package sizes, increasing the variation between the prices.

Table C1. Frequency of price changes in the CPI sample: matched pairs of comparable products (2016–2020)

| Category No. | Products subject to price control regulation | | Products not subject to price control regulation | |
|---|---|---|---|---|
| | No. of observations | Duration in months | No. of observations | Duration in months |
| 1. | 2,380 | 52.81 | 3,894 | 21.30 |
| 2. | 754 | 35.85 | 2,621 | 10.58 |
| 3. | 1,866 | 51.98 | 4,184 | 20.83 |
| 4. | 679 | 48.10 | 3,424 | 33.48 |
| 5. | 2,020 | 41.33 | 2,150 | 20.49 |
| 6. | 4,182 | 54.97 | 3,081 | 10.22 |
| 7. | 2,157 | 44.62 | 1,931 | 19.57 |
| 8. | 1,709 | 19.65 | 1,867 | 18.43 |
| 9. | 431 | 30.79 | 2,969 | 18.83 |
| 10. | 59 | 59.59 | 4,853 | 27.40 |

Notes: Products that are not subject to price controls are often sold in packages of different sizes than products that are subject to price controls.

Table C2. Descriptive statistics in the CPI sample: matched pairs of comparable products (2016–2020)

| Category No. | Products subject to price control regulation | | | Products not subject to price control regulation | | |
|---|---|---|---|---|---|---|
| | Mean | Std. | CV | Mean | Std. | CV |
| 1. | 4.67 | 0.1210 | 0.0259 | 5.77 | 0.3796 | 0.0658 |
| 2. | 12.51 | 0.2090 | 0.0501 | 18.89 | 4.3233 | 0.2289 |
| 3. | 1.47 | 0.1133 | 0.0771 | 5.56 | 0.7476 | 0.1345 |
| 4. | 7.70 | 0.0768 | 0.0199 | 4.79 | 0.4572 | 0.0954 |
| 5. | 2.22 | 0.0511 | 0.0230 | 10.22 | 2.8112 | 0.2751 |
| 6. | 5.72 | 0.2509 | 0.0439 | 13.10 | 2.5757 | 0.1966 |
| 7. | 11.39 | 0.0793 | 0.0070 | 20.37 | 4.4919 | 0.2205 |
| 8. | 5.08 | 0.0893 | 0.0176 | 9.10 | 2.9583 | 0.3251 |
| 9. | 7.00 | 0.2223 | 0.0318 | 15.34 | 3.1569 | 0.2058 |
| 10. | 2.05 | 0.1214 | | 5.26 | 1.7935 | |

<u>Notes</u>: Products that are not subject to price controls are often sold in packages of different sizes than products that are subject to price controls. The prices of two regulated products, semi-hard cheese – sold by weight (category 2) and butter (category 4) are given in the CBS data per 100 grams. To make the figures in the table comparable to those of the unregulated products, we report values representative of the median quantities that shoppers buy. We, therefore, report the prices for 300 grams for semi-hard cheese, and for 200 grams of butter.

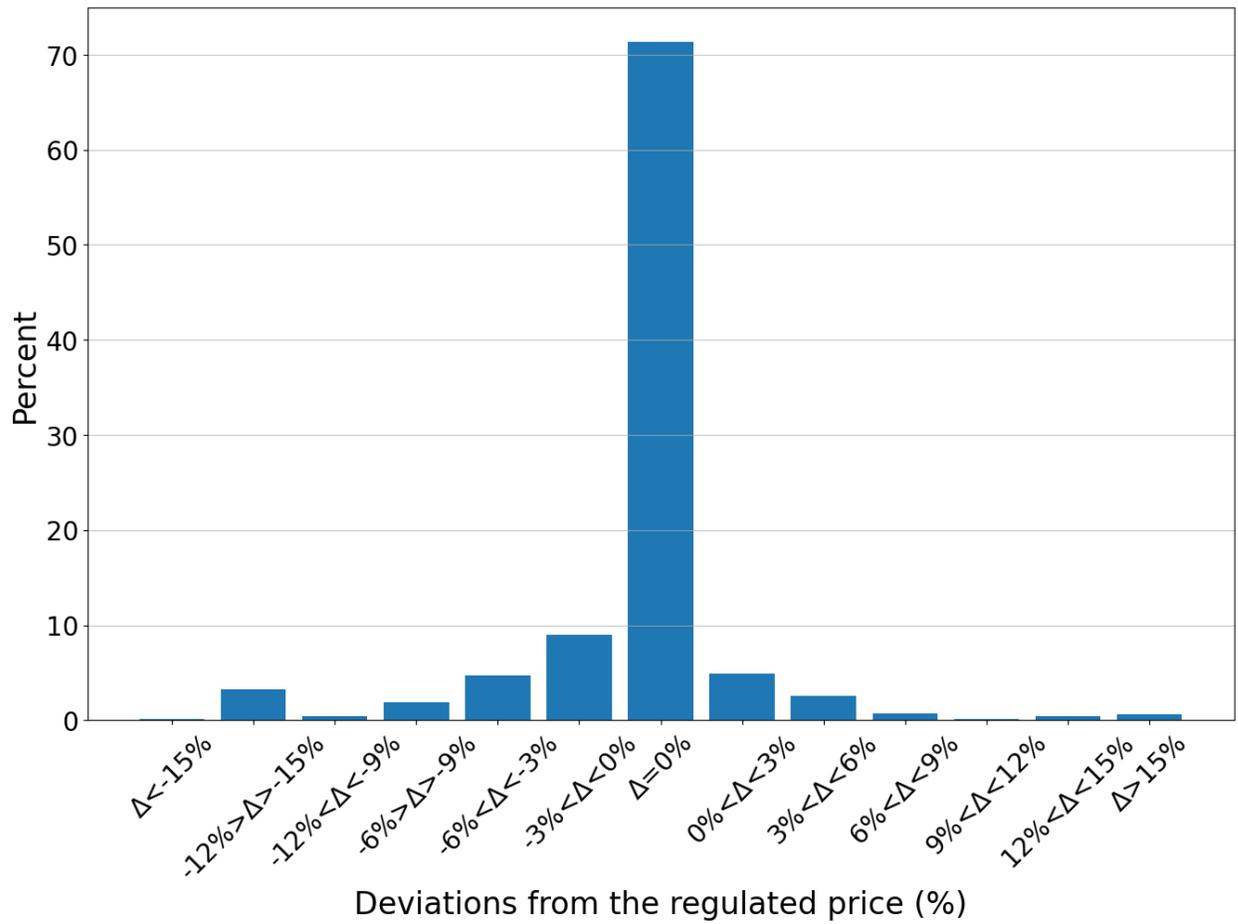

Figure C1. Deviations from the regulated price (2016–2020)

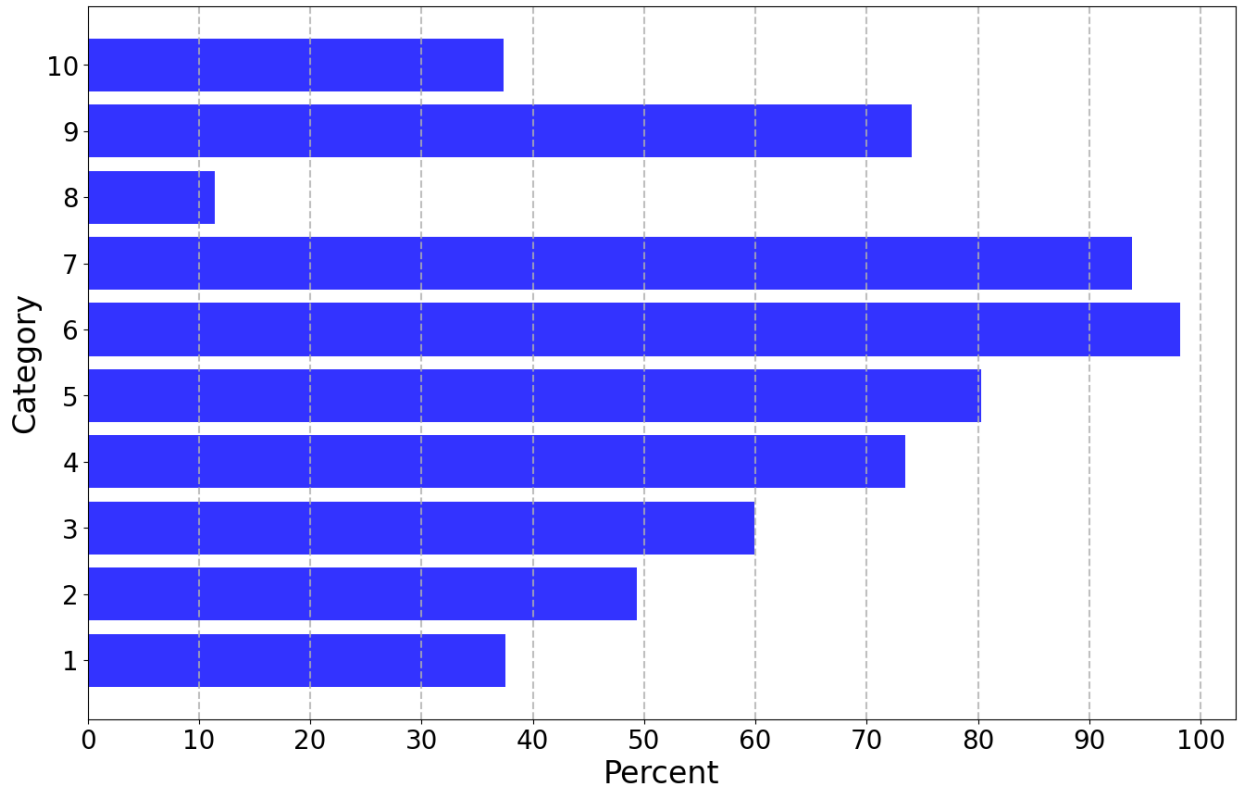

Figure C2. The percentage (of the number of units sold) of regulated products relative to comparable unregulated products in each category

Notes: Category 1 compares regulated soft cheese with non-regulated cottage cheese. Category 2 compares regulated semi-hard cheese with unregulated prepackaged semi-hard cheese. Category 3 compares regulated sour cream with unregulated plain yogurt. Category 4 compares regulated butter with unregulated margarine. Category 5 compares regulated whipped cream with unregulated non-dairy whipped cream. Category 6 compares regulated packaged milk with unregulated soy milk. Category 7 compares regulated eggs with unregulated cage-free/omega eggs. Category 8 compares regulated white bread – sliced with unregulated pita bread. Category 9 compares regulated whole wheat bread with unregulated light bread. Category 10 compares regulated salt with unregulated sugar. The bars depict the share of the regulated products in each category, in the total sales of the regulated and unregulated products. Blue (green) bars indicate that the share of the regulated products is smaller (larger) than 50%.

**Appendix D. Focusing on pairs of products with comparable prices**

In the paper, we report that shoppers recall the prices of unregulated products more precisely than the prices of regulated products. A possible explanation is that shoppers pay more attention to expensive products. They, therefore, might pay more attention to the prices of unregulated products because the prices of unregulated products tend to be higher than the prices of comparable regulated products.

To account for this possibility, we focus on two product categories in which the regulated and unregulated products have similar prices. There are products in categories 1 (soft cheese vs. cottage cheese) and 4 (butter vs. margarine). As can be seen in Appendix C, the prices of the regulated products in category 1 are similar. In Category 4, the price of butter (regulated) tends to be higher than the price of margarine (unregulated).

Table D1 reports the results of a set of regressions when we focus on these two categories. We report robust standard errors, clustered at the participants' level. The dependent variable in all regressions is the absolute value of the percentage error the shoppers make in recalling the prices. In column 1, the only covariate is a dummy for price control (1 = subject to price control, 0 = otherwise), in addition to supermarket and product-pair fixed effects. We also include random effects for participants. The estimated coefficient of the price control dummy is 0.036 and it is statistically significant at the 1% level.

In column 2, we add controls for the products' actual posted prices, the shopper's age, the household size, the number of cars the household possesses, the number of supermarkets the shoppers patronize, the average amount of money spent by the shopper on a shopping trip, and the proportion of the sampled 20 products which the shoppers correctly identified as regulated and unregulated. In addition, the regression includes dummy variables for gender (1 = woman, 0 = otherwise), marital status (1 = married, 0 = otherwise), academic degree (1 = with academic degree, 0 = otherwise), ultra-religious (1 = the shopper identifies himself/herself as ultra-religious, 0 = otherwise), frequent buyer (1 = shops more than once a week, 0 = otherwise), and for products with .90-ending prices.

The estimated coefficient of the price control dummy, 4.4%, is larger than in the previous estimation, and it remains statistically significant at the 1% level.

In column 3, we test the robustness of these results by adding fixed effects for the interactions between the supermarkets and the product pairs. These dummies capture systematic differences in the attention that shoppers who patronize different supermarkets pay to each product. We find that the coefficient of the regulated products is 4.9% and is statistically significant at the 1% level.

A possible explanation for our findings is that shoppers who buy regulated products do not buy unregulated ones. In that case, the differences we find might be due to differences between shoppers rather than the decisions that shoppers make regarding attention to the price information. In column 4, therefore, we account for this by estimating a fixed effects regression controlling for the interaction between consumers and product categories. The identification in this regression is based on shoppers who bought both regulated and unregulated products belonging to the same pairs (e.g., white cheese and cottage cheese). Using fixed effects regression leads to dropping all the coefficients that do not vary within participants. We find that the effect of the regulated prices remains positive ($\beta = 0.05$) and statistically significant.

It therefore seems that shoppers make larger errors when recalling the prices of regulated products even when the price level of the regulated and unregulated products is similar.

Table D1. Estimation results – Focusing on product categories with similar prices of regulated and unregulated products

| Variable | (1) | (2) | (3) | (4) |
|---|---|---|---|---|
| Price control (dummy) | 0.036*** | 0.044*** | 0.049*** | 0.050** |
|  | (0.010) | (0.013) | (0.013) | (0.024) |
| Product's posted price |  | 0.013 | 0.016 | 0.021 |
|  |  | (0.009) | (0.010) | (0.019) |
| Age |  | −0.000 | −0.000 |  |
|  |  | (0.000) | (0.000) |  |
| Household size |  | −0.003 | −0.003 |  |
|  |  | (0.003) | (0.003) |  |
| No. of cars |  | 0.006 | 0.009 |  |
|  |  | (0.008) | (0.008) |  |
| No. of supermarkets visited |  | −0.009 | −0.009 |  |
|  |  | (0.007) | (0.007) |  |
| Average amount spent |  | 0.007 | 0.006 |  |
|  |  | (0.005) | (0.005) |  |
| Woman (dummy) |  | −0.016 | −0.18* |  |
|  |  | (0.011) | (0.011) |  |
| Married (dummy) |  | −0.032** | −0.028** |  |
|  |  | (0.012) | (0.012) |  |
| Academic degree (dummy) |  | −0.004 | −0.004 |  |
|  |  | (0.012) | (0.011) |  |
| Ultra-religious (dummy) |  | −0.015 | −0.012 |  |
|  |  | (0.026) | (0.026) |  |
| Frequent shopper (dummy) |  | −0.015 | −0.016 |  |
|  |  | (0.020) | (0.015) |  |
| Percentage of correct recalls of regulated prices |  | −0.043 | −0.045 |  |
|  |  | (0.074) | (0.073) |  |
| .90-ending price (dummy) |  | 0.006 | −0.002 | −0.041 |
|  |  | (0.020) | (0.018) | (0.019) |
| Constant | 0.17*** | 0.163** | 0.131 | −0.079 |
|  | (0.033) | (0.078) | (0.081) | (0.325) |
| $R^2$ | 0.23 | 0.25 | 0.29 | 0.02 |
| N | 1,021 | 1,021 | 1,021 | 1,021 |

Notes:
1. In all columns, the dependent variable is $ln\left|\frac{Recall.price_{i,j,t}}{Actual.price_{j,t}}\right|$, where $i$ stands for the participant, $j$ for a product, and $t$ for the week of the observation.
2. Columns 1–3 report the results of regressions with random effects for participants. Column 4 reports the regression results with fixed effects for participants×product pairs.
4. The standard errors are robust and clustered at the participants' level.
5. All the regressions include product-pair and supermarket dummies (not reported to save space).
6. Observations with relative errors above 100% were excluded from the regressions.
7. Including observations only of products from categories 1 and 4.
8. ***, **, and * indicate significance at 1%, 5%, and 10%, respectively.

**Appendix E. Using the levels of the price recall errors**

In the paper, we report that shoppers make larger errors when recalling the prices of regulated products than when recalling the prices of unregulated products. A possible concern is that these results might be driven by our measuring errors in relative terms.

For example, assume that shoppers make errors of similar size (in NIS) regardless of the price of a product. In that case, because regulated products tend to have lower prices than unregulated products, using relative terms would lead to the wrong conclusion that shoppers make larger errors when recalling the prices of regulated products than when recalling the prices of unregulated products.

To account for this possibility, we re-estimate the regressions that we estimate in the paper, using the level of the price recall errors as the dependent variables, rather than the absolute value of the relative recall errors. Table E1 reports the results.

In column 1, the only covariate is a dummy for price control (1 = subject to price control, 0 = otherwise), in addition to supermarket and product-pair fixed effects. We also include random effects for participants. The estimated coefficient of the price control dummy is statistically significant at the 1% level, and its magnitude suggests that shoppers make errors that are NIS 0.546 larger in recalling the prices of regulated products than when recalling the prices of unregulated products.

In column 2, we add controls for the products' actual posted prices, the shopper's age, the household size, the number of cars the household possesses, the number of supermarkets the shoppers patronize, the average amount of money spent on a shopping trip, and the proportion of the sampled 20 products which the shoppers correctly identified as regulated and unregulated. In addition, the regression includes dummy variables for gender (1 = woman, 0 = otherwise), marital status (1 = married, 0 = otherwise), academic degree (1 = with academic degree, 0 = otherwise), ultra-religious (1 = the shopper identifies himself/herself as ultra-religious, 0 = otherwise), frequent buyer (1 = shops more than once a week, 0 = otherwise), and for products with .90-ending prices.

The estimated coefficient of the price control dummy is negative, –0.254, but it is not statistically significant. In column 3, we test the robustness of these results by adding fixed effects for the interactions between the supermarkets and the product pairs. These dummies capture systematic differences in the attention that shoppers who patronize different supermarkets pay to each product. We find that the coefficient of the price control dummy is 0.132 and is not statistically significant.

A possible explanation for our findings is that shoppers who buy regulated products do not buy unregulated ones. In that case, the differences we find might be due to differences between shoppers rather than differences in their attention to the price information. In column 4, we account for this possibility by estimating a fixed effects regression controlling for the interaction between consumers and product categories. The identification in this regression is based on shoppers who bought both regulated and unregulated products belonging to the same pairs (e.g., white cheese and cottage cheese). Using fixed effects regression leads to dropping all the coefficients that do not vary within participants. We find that the coefficient of the price control dummy is 0.337 and statistically significant at the 5% level.

We conclude that even if we measure the size of the errors in levels rather than in relative terms, the shoppers still tend to make larger errors when recalling the prices of regulated prices.

Table E1. Estimation results – levels of the price recall error

| Variable | (1) | (2) | (3) | (4) |
|---|---|---|---|---|
| Price control (dummy) | 0.546*** (0.100) | −0.254 (0.157) | 0.132 (0.111) | 0.337** (0.154) |
| Product's posted price | | −0.205*** (0.027) | −0.094*** (0.027) | -0.063 (0.052) |
| Age | | −0.007** (0.004) | −0.008 (0.004) | |
| Household size | | −0.019 (0.029) | −0.026 (0.028) | |
| No. of cars | | 0.020 (0.071) | 0.016 (0.070) | |
| No. of supermarkets visited | | −0.083 (0.063) | -0.085 (0.060) | |
| Average amount spent | | 0.146*** (0.046) | 0.138*** (0.0469) | |
| Woman (dummy) | | −0.044 (0.090) | −0.039 (0.087) | |
| Married (dummy) | | −0.217** (0.106) | −0.255** (0.102) | |
| Academic degree (dummy) | | −0.094 (0.098) | −0.064 (0.098) | |
| Ultra-religious (dummy) | | −0.028 (0.311) | −0.003 (0.279) | |
| Frequent shopper (dummy) | | −0.105 (0.122) | -0.099 (0.130) | |
| Percentage of correct recalls of regulated prices | | −0.094 (0.098) | −0.404 (0.557) | |
| .90-ending price (dummy) | | 0.019 (0.147) | −0.174 (0.178) | −0.487** (0.242) |
| Constant | 0.170 (0.323) | 2.022*** (0.504) | 1.979*** (0.669) | −2.647 (2.998) |
| $R^2$ | 0.06 | 0.13 | 0.27 | 0.01 |
| N | 3,162 | 3,162 | 3,162 | 3,162 |

Notes:
1. In all columns, the dependent variable is the level of the price recall error.
2. Columns 1–3 report the results of regressions with random effects for participants. Column 4 reports the regression results with fixed effects for participants×product pairs.
4. The standard errors are robust, and clustered at the participants' level.
5. All the regressions include product-pair and supermarket dummies (not reported to save space).
7. Using data from the pooled sample.
7. ***, **, and * indicate significance at 1%, 5%, and 10%, respectively.

**Appendix F. Excluding from the analysis the soft cheese - cottage cheese pair**

The pair of soft cheese (regulated) and cottage cheese (unregulated) is unique because, as discussed in Hendel et al. (2017), the price of cottage cheese had risen by 43% between 2006 and 2011. This led to widespread calls to boycott cottage cheese until retailers brought the price down. The cottage cheese boycott later expanded, with tens of thousands of consumers going to the streets to protest the high cost of living.

Hendel et al. (2017) note that the price of cottage cheese remained low for an extended period after the boycott, which underscores the public interest in its price. It is worthwhile, therefore, to test whether our results are robust to excluding the pair of soft cheese – cottage cheese. By excluding these products, we can test whether our results are driven by some special interest that shoppers have in these particular products.

We, therefore, re-estimate regressions similar to the ones reported in Table 3 in the paper, after we have excluded the observations on the soft cheese – cottage cheese pair. The results are reported in Table F1.

In column 1, the only covariate is a dummy for price control (1 = subject to price control, 0 = otherwise), in addition to supermarket and product-pair fixed effects. We also include random effects for participants. The estimated coefficient of the price control dummy is statistically significant at the 1% level, and its magnitude suggests that shoppers make errors that are 3.6% larger in recalling the prices of regulated products than when recalling the prices of unregulated products. This amounts to about 13.2% of the average absolute value of the price recall error of 27.3%.

In column 2, we add controls for the products' actual posted prices, the shopper's age, the household size, the number of cars the household possesses, the number of supermarkets the shoppers patronize, the average amount of money spent by the shopper on a shopping trip, and the proportion of the sampled 20 products which the shoppers correctly identified as regulated and unregulated. In addition, the regression includes dummy variables for gender (1 = woman, 0 = otherwise), marital status (1 = married, 0 = otherwise), academic degree (1 = with academic degree, 0 = otherwise), ultra-religious (1 = the shopper identifies himself/herself as ultra-religious, 0 = otherwise), frequent buyer (1 = shops more than once a week, 0 = otherwise), and for products with .90-ending prices.

The estimated coefficient of the price control dummy, 4.9%, is larger than in the previous regression, and it remains statistically significant at the 1% level.

In column 3, we test the robustness of these results by adding fixed effects for the interactions between the supermarkets and the product pairs, which capture systematic differences in the attention that shoppers who patronize different supermarkets pay to each product. We find that the coefficient of the regulated products is 4.3%.

A possible explanation for our findings is that shoppers who buy regulated products do not buy unregulated ones. In that case, the differences we find might be due to differences between shoppers rather than to the decisions that shoppers make regarding attention to the price information. In column 4, therefore, we control for this by estimating a fixed effects regression controlling for the interaction between consumers and product categories. The identification in this regression is based on shoppers who bought both the regulated and unregulated products belonging to the same pairs. Using fixed effects regression leads to dropping all the coefficients that do not vary within participants. We find that the effect of the regulated prices remains positive ($β = 0.095$) and statistically significant.

We conclude that excluding observations on soft cheese – cottage cheese pair, does not alter our main results.

Table F1. Estimation results – after excluding the soft cheese - cottage cheese pair

| Variable | (1) | (2) | (3) | (4) |
|---|---|---|---|---|
| Price control (dummy) | 0.036*** | 0.049*** | 0.043*** | 0.095** |
|  | (0.012) | (0.015) | (0.164) | (0.043) |
| Product's posted price |  | 0.003* | 0.003* | 0.001 |
|  |  | (0.002) | (0.002) | (0.004) |
| Age |  | −0.001*** | −0.001*** |  |
|  |  | (0.000) | (0.000) |  |
| Household size |  | −0.001 | −0.001 |  |
|  |  | (0.003) | (0.003) |  |
| No. of cars |  | 0.005 | 0.007 |  |
|  |  | (0.006) | (0.006) |  |
| No. of supermarkets visited |  | −0.006 | −0.006 |  |
|  |  | (0.006) | (0.006) |  |
| Average amount spent |  | 0.001 | −0.000 |  |
|  |  | (0.004) | (0.004) |  |
| Woman (dummy) |  | −0.008 | −0.005 |  |
|  |  | (0.009) | (0.009) |  |
| Married (dummy) |  | −0.008 | −0.005 |  |
|  |  | (0.010) | (0.010) |  |
| Academic degree (dummy) |  | −0.011 | −0.012 |  |
|  |  | (0.009) | (0.009) |  |
| Ultra-religious (dummy) |  | −0.006 | 0.007 |  |
|  |  | (0.025) | (0.025) |  |
| Frequent shopper (dummy) |  | −0.006 | −0.006 |  |
|  |  | (0.012) | (0.012) |  |
| Percentage of correct recalls of regulated prices |  | −0.113** | −0.061 |  |
|  |  | (0.055) | (0.056) |  |
| .90-ending price (dummy) |  | 0.002 | −0.009 | −0.060 |
|  |  | (0.012) | (0.016) | (0.041) |
| Constant | 0.424*** | 0.437*** | 0.732*** | −1.743** |
|  | (0.031) | (0.062) | (0.046) | (0.687) |
| $R^2$ | 0.24 | 0.24 | 0.33 | 0.01 |
| $N$ | 2,488 | 2,488 | 2,488 | 2,488 |

Notes:

1. In all columns, the dependent variable is $ln\left|\frac{Recall.price_{i,j,t}}{Actual.price_{j,t}}\right|$, where $i$ stands for the participant, $j$ for a product, and $t$ for the week of the observation.
2. Columns 1–3 report the results of regressions with random effects for participants. Column 4 reports the regression results with fixed effects for participants×product pairs.
4. The standard errors are robust and clustered at the participants' level.
5. All the regressions include product-pair and supermarket dummies (not reported to save space).
6. Observations with relative errors above 100% were excluded from the regressions.
7. ***, **, and * indicate significance at 1%, 5%, and 10%, respectively.

**Appendix G. A translation of the survey questionnaire**

Out of the following products, please refer only to the ones that you purchased during the current visit to the supermarket. If you did not purchase a certain product, please continue to the next one on the list.

1. **Cottage cheese, 250 grams:**
   Brand: _________________
   Price: _________________

2. **White cheese, 5% fat, 250 grams:**
   Brand: _________________
   Price: _________________

3. **Hard ("yellow") cheese, 28% fat ("Emek" or "Noam"), by weight:**
   Brand: _________________
   Price: _________________

4. **Hard ("yellow") cheese, 28% fat ("Emek" or "Noam"), packaged (Emek – 400 grams, Noam – 360/500 grams):**
   Brand: _________________
   Price: _________________

5. **"Gil" or "Eshel" sour milk, 200 grams:**
   Brand: _________________
   Price: _________________

6. **Yogurt (natural flavor):**
   Brand: _________________
   Price: _________________

7. **Butter 100/200 grams:**

    Brand: _________________

    Price: _________________

8. **Margarine, 200 grams:**

    Brand: _________________

    Price: _________________

9. **Whip cream, 38% fat, 250 grams:**

    Brand: _________________

    Price: _________________

10. **Non-dairy cream:**

    Brand: _________________

    Price: _________________

11. **Milk, 3% fat, 1 litter in a cardboard package:**

    Brand: _________________

    Price: _________________

12. **Soy milk, 1 litter:**

    Brand: _________________

    Price: _________________

13. **12 Large eggs, in cardboard package:**

    Brand: _________________

    Price: _________________

14. **Large extra fresh, "omega enriched", or "freedom" eggs:**

    Brand: ________________

    Price: ________________

15. **Standard bread, not sliced, white/whole-wheat:**

    Brand: ________________

    Price: ________________

16. **Standard bread, sliced, white/whole-wheat (white – 500 grams, whole-wheat – 750 grams)**:

    Brand: ________________

    Price: ________________

17. **A package of pita bread (5/10 units):**

    Brand: ________________

    Price: ________________

18. **Low carb. Bread, sliced:**

    Brand: ________________

    Price: ________________

19. **Table salt, paper bag, 1 kg:**

    Brand: ________________

    Price: ________________

20. **White sugar, paper bag, 1 kg:**

    Brand: ________________

    Price: ________________

**Socio-demographic information**

1. **Gender:**

   a. Male    b. Female

2. **Marital status:**

   a. Bachelor   b. Married    c. Divorced    d. Widower

   e. Other: ______

3. **Age:** ______

4. **Education:**

   a. Primary school   b. High school    c. Academic    d. Professional

   e. Other: ______

5. **Number of family members:** ____________

6. **How would you define your religious affiliation:**

   a. Ultra-religious    b. Religious    c. Traditional  d. Secular    e. Other: _______

7. **In how many food stores do you shop on a regular basis?**

   a. 0   b. 1    c. 2    d. 3 or more

8. **How frequently do you shop in this store?**

   a. More than once a week   b. Once a week

   c. Once every 2 weeks     d. Infrequently

9. **How much money do you typically spend when shopping in this store?**

   a. Less than NIS 100     b. NIS 100 – 250     c. NIS 400 – 550

   d. NIS 550 – 700         e. More than NIS 700

The prices of which of the following products are regulated by the government?

1. Cottage cheese:                                    yes/no
2. White cheese:                                      yes/no
3. Hard ("yellow") cheese, "Emek", by weight:         yes/no
4. Hard ("yellow") cheese, "Emek", packaged:          yes/no
5. "Eshel" sour milk:                                 yes/no
6. Yogurt, natural flavor:                            yes/no
7. Butter:                                            yes/no
8. Margarine:                                         yes/no
9. Whip cream (38% fat):                              yes/no
10. Non-dairy cream:                                  yes/no
11. Milk, 3%, cardboard package:                      yes/no
12. Soy milk:                                         yes/no
13. 12 Eggs, cardboard package:                       yes/no
14. "Freedom" eggs:                                   yes/no
15. Standard bread, white, not sliced:                yes/no
16. Standard bread, white, sliced:                    yes/no
17. 10 pita bread:                                    yes/no
18. Low carb. white bread:                            yes/no
19. Table salt, 1 kg:                                 yes/no
20. White sugar, 1 kg:                                yes/no

**References**

Ater, I., and O. Gerlitz (2017), "Round Prices and Price Rigidity: Evidence from Outlawing Odd Prices," ***Journal of Economic Behavior and Organization*** 144(C), 188–203.

Hendel, I., Lach, S., and Spiegel, Y. (2017), "Consumers' Activism: The Cottage Cheese Boycott," ***Rand Journal of Economics*** 48(4), 972–1003.

Sayag, D., A. Snir and D. Levy (2024), "Price Setting Rules, Rounding Tax, and Inattention Penalty," Working Paper.

Snir, A., D. Levy and H. Chen (2017), "End of 9-Endings, Price Recall, and Price Perceptions," ***Economic Letters*** 155, 157–163.